\documentclass[11pt]{article}
\usepackage[utf8]{inputenc}

\usepackage[paperwidth=8.5in,paperheight=11in,portrait,top=1in,bottom=1.in,left=1.15in,right=1.15in]{geometry}
\usepackage{authblk}

\usepackage{amsfonts,amsmath,amssymb,amsthm}
\allowdisplaybreaks
\usepackage{latexsym,mathrsfs,mathtools,bm}
\usepackage{braket}
\usepackage{pgfplots}
\usepackage{xcolor}
\usepackage{graphicx,subcaption,epsfig,caption,float,xcolor}
\usepackage{enumitem}

\usepackage[hidelinks]{hyperref}
\usepackage{bookmark}

\theoremstyle{plain}
\newtheorem{thm}{Theorem}[section]

\newtheorem{rem}[thm]{Remark}

\numberwithin{equation}{section}

\def\cA{{\mathcal A}}      \def\cC{{\mathcal C}}
      
   \def\cH{{\mathcal H}}   
      
   \def\cN{{\mathcal N}}   \def\cO{{\mathcal O}}
      
\def\cS{{\mathcal S}}      
   \def\cW{{\mathcal W}}

\title{\bf A discrete Smorodinsky--Winternitz I superintegrable system}

\renewcommand*{\Affilfont}{\normalsize\small}
\author[1]{Vutha Vichhea Chea}
\author[2]{Luc Vinet}
\affil[1,2]{Centre de Recherches Math\'ematiques, Universit\'e de Montr\'eal, P.O. Box 6128, Centre-ville Station, Montr\'eal (Qu\'ebec), H3C 3J7, Canada. \vspace{.5em}}

{
	\makeatletter
	\renewcommand\AB@affilsepx{: \protect\Affilfont}
	\makeatother
	\affil[ ]{E-mail addresses}
	\makeatletter
	\renewcommand\AB@affilsepx{, \protect\Affilfont}
	\makeatother
	\affil[1]{vutha.vichhea.chea@umontreal.ca}
	\affil[2]{luc.vinet@umontreal.ca}
}

\begin{document}

\maketitle

\begin{abstract}

We construct a finite discrete realization of the
Smorodinsky--Winternitz I superintegrable system on a triangular region of the two-dimensional square lattice. The construction is based on a pair of commuting number operators with finite spectrum together with an associated ladder-operator structure. We show that the resulting model is
maximally superintegrable and that its symmetry algebra admits a
Hahn-algebra presentation. Its spectral problem is solved exactly in terms of the bivariate dual Hahn polynomials of Tratnik type. Finally, we show that the continuum limit recovers the continuous Smorodinsky--Winternitz I system together with its ladder operators, eigenfunctions and symmetry algebra, thereby establishing the present construction as a genuine finite discrete realization of the continuous model.

\end{abstract}

\section{Introduction}

A quantum system with Hamiltonian $H$ in $n$ dimensions is said to be \emph{superintegrable} \cite{Mil-Pos-Win-13} if it admits $n+k$, where $1\le k\le n-1$, algebraically independent Hermitian operators $C_{i \in \{1,\dots,n+k\}}$ that commute with $H$. They are labeled as symmetry operators or integrals of motion. The symmetry operators are \emph{algebraically independent} if no polynomials formed entirely out of $C_i$ using the symmetrized Jordan product vanish identically. If $k=n-1$, the system is said to be \emph{maximally superintegrable}. Beyond their intrinsic interest, maximally superintegrable systems provide controlled settings for the study of hidden symmetries and their algebraic structures.

In two dimensions, superintegrable systems with second-order integrals
of motion were classified by Winternitz, Smorodinsky and collaborators
\cite{frivs1965higher,winternitz1966symmetry}. These systems were shown
to possess polynomial symmetry algebras and to be exactly solvable
\cite{granovskii1992mutual,letourneau1995superintegrable,Das-01,Mar-09}.
The close relation between maximal superintegrability and exact
solvability was emphasized in \cite{Tem-Tur-Win-00}. The algebraic
description of superintegrable systems has been investigated
extensively, notably through the appearance of Racah-type and related
algebras \cite{kalnins2007wilson,genest2014superintegrability}.
In the context of studies on multivariate orthogonal polynomials, difference equations with degeneracies understood as stemming from symmetries have been presented in
\cite{Ata-Pog-Vic-Wol-01,Ata-Nat-Pog-Geo-Vic-Wol-01,de2017superintegrable,Gen-Vin-14,iliev2018symmetry,Ili-Xu-20,kalnins2011two,Sas-23} for instance.
More recently, superintegrability has been revisited from a
representation-theoretic viewpoint
\cite{reshetikhin2015degenerately,arthamonov2021superintegrable}.
The list of references cited here is necessarily partial.

Discrete analogues of superintegrable quantum systems have also been
investigated. A paradigmatic example is the finite discrete two-dimensional
oscillator introduced in \cite{Mik-Pos-Vin-Zhe-2012}, based on the
bivariate Krawtchouk polynomials of Griffiths type \cite{Gri-71}. That model
provides a discretization of the isotropic harmonic oscillator and
remarkably possesses the same $\mathfrak{su}(2)$ symmetry algebra as
its continuous counterpart.

The present work is part of a broader program devoted to the
construction of finite discrete superintegrable quantum systems \cite{Gab-Gen-Lem-Vin-15,Gen-Mik-Vin-Yu-17,Mik-Pos-Vin-Zhe-2012}. Beyond the
finite discrete oscillator model based on bivariate Krawtchouk polynomials, it
is natural to ask whether other members of the
Smorodinsky--Winternitz family admit finite discrete realizations preserving
maximal superintegrability and exact solvability. The answer turns out
to be affirmative.

In the present paper we construct a finite realization of the
Smorodinsky--Winternitz I system on a triangular region of the two-dimensional square lattice. As will be shown, the model is maximally superintegrable,
its symmetry algebra admits a Hahn-algebra presentation \cite{granovskii1992mutual,Fra-Gab-Vin-Vin-Zhe-19}, and its
spectral problem is solved exactly in terms of the associated
bivariate dual Hahn polynomials of Tratnik type. We further show that an appropriate
continuum limit reproduces the continuous
Smorodinsky--Winternitz I system \cite{frivs1965higher,winternitz1966symmetry} together with its Laguerre--Laguerre eigenfunctions and its symmetry algebra which also admits a Hahn-algebra presentation.
A companion paper is devoted to the discretization of the
Smorodinsky--Winternitz II system \cite{Ber-Che-Vin-26}. Together, these constructions
provide further evidence that maximal superintegrability, exact
solvability and the underlying algebraic structures can survive
discretization on finite lattices.

The paper is organized as follows. In
Section~\ref{sec:model-definition}, we define the discrete model and
its Hamiltonian. In Section~\ref{sec:supint-exact-solv}, we construct the associated ladder operators to establish the dynamical algebra,
determine the symmetry algebra, and solve the spectral
problem. Section~\ref{sec:continuum-limit} is devoted to the continuum
limit of the discrete model. In Section~\ref{sec:conc}, we discuss the results and present some future outlooks. Appendix~\ref{app:ladder-coefficients} entails the explicit coefficients of the ladder operators. Finally, Appendices \ref{app:hyp-poly} and \ref{app:Tratnik-poly} detail some properties and calculations of the model.

\section{A discrete Smorodinsky--Winternitz I model}
\label{sec:model-definition}

\begin{figure}[htbp]
\centering
\begin{tikzpicture}[scale=1.1]

\def\N{5}

\draw[gray!30, step=1] (0,0) grid (\N+0.5,\N+0.5);

\draw[->, thick] (0,0) -- (\N+0.8,0) node[right] {$x_1$};
\draw[->, thick] (0,0) -- (0,\N+0.8) node[above] {$x_2$};

\draw (0,0) -- (0,-0.1) node[below] {$0$};
\draw (0,0) -- (-0.1,0) node[left] {$0$};

\draw (\N,0) -- (\N,-0.1) node[below] {$N$};
\draw (0,\N) -- (-0.1,\N) node[left] {$N$};

\foreach \x in {0,...,\N}
{
    \pgfmathtruncatemacro{\ymax}{\N-\x}
    \foreach \y in {0,...,\ymax}
    {
        \fill (\x,\y) circle (2pt);
    }
}

\end{tikzpicture}
\caption{The discrete position space $\mathfrak{R}(N)$ of \eqref{eq:triangular-region},
an isosceles right triangular region of side length $N$ in the first quadrant of the
$(x_1,x_2)$-plane (here $N=5$).}
\label{fig:triangular-region}
\end{figure}
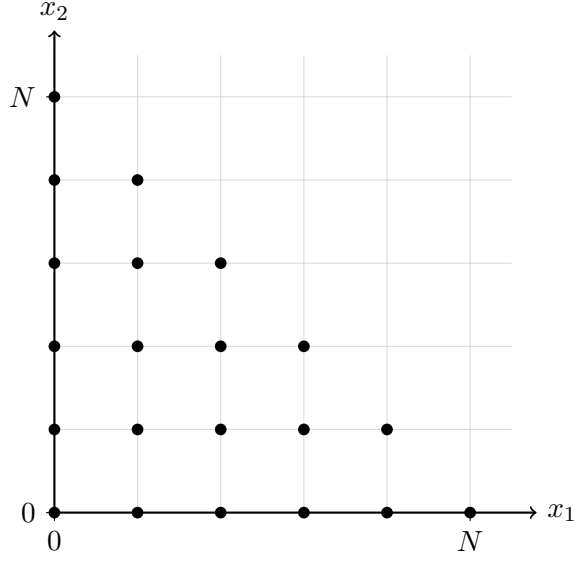

We introduce a discrete model defined on a triangular region of the two-dimensional square lattice. The position space is the region (see Figure~\ref{fig:triangular-region})
\begin{equation}
    \mathfrak{R}(N)
    = \bigl\{ (x_1,x_2) \in \mathbb{N}_0^2 \;\big|\; 0 \le x_1 + x_2 \le N \bigr\},
    \qquad N \in \mathbb{N},
    \label{eq:triangular-region}
\end{equation}
adapted to the dual Hahn structure that will appear in the spectral problem. The corresponding Hilbert space is defined as the space generated by the set of position eigenstates localized in $\mathfrak{R}(N)$,
\begin{equation}
    \mathfrak{H} = \text{Span}\{\ket{x_1,x_2} | \, (x_1,x_2) \in \mathfrak{R}(N) \}.
\end{equation}
The inner product/orthogonality relation is fixed by
\begin{equation}
    \braket{x_1,x_2|y_1,y_2} =
    \frac{\delta_{x_1,y_1}\delta_{x_2,y_2}}{w(x_1,x_2)},
    \label{eq:ortho-rel-position}
\end{equation}
where $\delta_{x_i,y_i}$ is the Kronecker delta and $w(x_1,x_2)$ is a positive local weight function to be determined shortly. The completeness relation
\begin{equation}
    \sum_{(x_1,x_2) \in \mathfrak{R}(N)} w(x_1,x_2)\ket{x_1,x_2}\bra{x_1,x_2} = 1
    \label{eq:completeness-relation}
\end{equation}
ensures that any vector state $\ket{\psi} \in \mathfrak{H}$ admits the expansion
\begin{equation}
    \ket{\psi} =
    \sum_{(x_1,x_2) \in \mathfrak{R}(N)} w(x_1,x_2)\psi(x_1,x_2)\ket{x_1,x_2},
    \qquad
    \braket{x_1,x_2|\psi} = \psi(x_1,x_2).
    \label{eq:state-expansion}
\end{equation}
For a linear operator $\cO$, its action on $\ket{\psi}$ is given by
\begin{equation}
    \cO\ket{\psi} \quad = \sum_{(x_1,x_2) \in \mathfrak{R}(N)} w(x_1,x_2)\cO\psi(x_1,x_2)\ket{x_1,x_2}, \qquad \cO\psi(x_1,x_2) = \bra{x_1,x_2}\cO\ket{\psi}.
    \label{eq:linear-operator-action}
\end{equation}
The Hermitian adjoint $\cO^\dagger$ is defined by the relation
\begin{equation}
    \sum_{(x_1,x_2) \in \mathfrak{R}(N)} w(x_1,x_2)\phi^*(x_1,x_2)\cO\psi(x_1,x_2)  =
    \sum_{(x_1,x_2) \in \mathfrak{R}(N)} w(x_1,x_2)\cO^\dagger\phi^*(x_1,x_2)\psi(x_1,x_2),
    \label{eq:adjoint-rule}
\end{equation}

\subsection{Hamiltonian operator}

The Hamiltonian operator is taken to be the sum of two commuting number operators,
\begin{equation}
    H = H_1 + H_2,
    \qquad H_i = N_i + \frac{\alpha_i + 1}{2},
    \qquad [N_i , N_j] = 0, \qquad i,j = 1,2,
    \label{eq:hamiltonian-operator}
\end{equation}
where $\alpha_i > -1$ are fixed parameters. The expressions for $N_i$ are chosen so that these operators preserve the space of functions on $\mathfrak{R}(N)$, are self-adjoint with respect to a local positive weight function and admit a common family of polynomial eigenfunctions. Explicitly,
\begin{align}
    N_1 = \sum_{\vec{\eta} \in S_1}N_1^{\eta_1,\eta_2}\Delta_{\eta_1,\eta_2}, \qquad N_2 = \sum_{\vec{\eta} \in S_2}N_2^{\eta_1,\eta_2}\Delta_{\eta_1,\eta_2} - N_1.
    \label{eq:number-operators}
\end{align}
The coefficients are given by
\begin{align}
    S_1 =&\ \{(1,-1),(-1,1)\},
    \nonumber\\
    N^{1,-1}_1 =& -\frac{x_2 \left(\alpha _2+2 N-2 x_1-x_2+1\right) \left(\alpha _1+\alpha _2+2 N-x_1+2\right)}{\left(\alpha _1+\alpha _2+2 N-2 x_1+1\right)_2},
   \nonumber\\
    N^{-1,1}_1 =& -\frac{x_1 \left(\alpha _1+x_2+1\right) \left(\alpha _1+\alpha _2+2 N-2 x_1-x_2+2\right)}{\left(\alpha _1+\alpha _2+2 N-2 x_1+2\right)_2},
\end{align}
\begin{align}
    S_2 =&\ \{(-1,2),(1,0),(0,1),(1,-1),(-1,1),(-1,0),(0,-1),(1,-2)\},
    \nonumber\\
    N_2^{-1,2} =& -\frac{x_1 \left(N-x_1-x_2\right){}^2 \left(\alpha _1+x_2+1\right)_2}{\left(\alpha _2+2 N-2 x_1-2 x_2\right)_2 \left(\alpha _1+\alpha _2+2 N-2 x_1+2\right)_2},
    \nonumber\\
    N_2^{1,0} =& -\frac{\left(N-x_1-x_2\right){}^2 \left(\alpha _2+2 N-2 x_1-x_2\right)_2 \left(\alpha _1+\alpha _2+2 N-x_1+2\right)}{\left(\alpha _2+2 N-2 x_1-2 x_2\right)_2 \left(\alpha _1+\alpha _2+2 N-2 x_1+1\right)_2},
    \nonumber\\
    N_2^{0,1} =& -\frac{\left(\alpha _1+\alpha _2+2 N+3\right) \left(N-x_1-x_2\right){}^2 \left(\alpha _1+x_2+1\right) \left(\alpha _2+2 N-2 x_1-x_2+1\right)}{\left(\alpha _2+2 N-2 x_1-2 x_2\right)_2
   \left(\alpha _1+\alpha _2+2 N-2 x_1+1\right) \left(\alpha _1+\alpha _2+2 N-2 x_1+3\right)},
   \nonumber\\
   N_2^{1,-1} =& -\left(\alpha _2 \left(2 N-2 x_1-2 x_2+1\right)+2 \left(N-x_1-x_2\right)_2\right)
   \nonumber\\
   &\times\frac{x_2 \left(\alpha _2+2 N-2 x_1-x_2+1\right) \left(\alpha _1+\alpha _2+2 N-x_1+2\right)}{\left(\alpha _2+2 N-2
   x_1-2 x_2\right) \left(\alpha _2+2 N-2 x_1-2 x_2+2\right) \left(\alpha _1+\alpha _2+2 N-2 x_1+1\right)_2},
   \nonumber\\
   N_2^{-1,1} =& -\left(\alpha _2 \left(2 N-2 x_1-2 x_2+1\right)+2 \left(N-x_1-x_2\right)_2\right)
   \nonumber\\
   &\times\frac{x_1 \left(\alpha _1+x_2+1\right) \left(\alpha _1+\alpha _2+2 N-2 x_1-x_2+2\right)}{\left(\alpha _2+2 N-2 x_1-2
   x_2\right) \left(\alpha _2+2 N-2 x_1-2 x_2+2\right) \left(\alpha _1+\alpha _2+2 N-2 x_1+2\right)_2},
   \nonumber\\
   N_2^{-1,0} =& -\frac{x_1 \left(\alpha _2+N-x_1-x_2+1\right){}^2 \left(\alpha _1+\alpha _2+2 N-2 x_1-x_2+2\right)_2}{\left(\alpha _2+2 N-2 x_1-2 x_2+1\right)_2 \left(\alpha _1+\alpha _2+2 N-2 x_1+2\right)_2},
   \nonumber\\
   N_2^{0,-1} =& -\frac{x_2 \left(\alpha _1+\alpha _2+2 N+3\right) \left(\alpha _2+N-x_1-x_2+1\right){}^2 \left(\alpha _1+\alpha _2+2 N-2 x_1-x_2+2\right)}{\left(\alpha _2+2 N-2 x_1-2 x_2+1\right)_2
   \left(\alpha _1+\alpha _2+2 N-2 x_1+1\right) \left(\alpha _1+\alpha _2+2 N-2 x_1+3\right)},
   \nonumber\\
   N_2^{1,-2} =& -\frac{\left(x_2-1\right)_2 \left(\alpha _2+N-x_1-x_2+1\right){}^2 \left(\alpha _1+\alpha _2+2 N-x_1+2\right)}{\left(\alpha _2+2 N-2 x_1-2 x_2+1\right)_2 \left(\alpha _1+\alpha _2+2
   N-2 x_1+1\right)_2},
\end{align}
where $(a)_k = a(a+1)\cdots(a+k-1)$ is the Pochhammer symbol and $\Delta_{\eta_1, \eta_2}$, where $(\eta_1,\eta_2)\in\mathbb{Z}^2$, denotes a finite difference operator whose action on a function $f$, reads as
\begin{align}
    \Delta_{\eta_1,\eta_2}f(x_1,x_2)
    =
    \begin{cases}
        f(x_1+\eta_1, x_2+\eta_2) - f(x_1,x_2) \qquad \text{if} \qquad (\eta_1,\eta_2) \ne (0,0),
        \\
        f(x_1,x_2) \qquad \text{if} \qquad (\eta_1,\eta_2) = (0,0).
    \end{cases}
    \label{eq:difference-operator}
\end{align}

The weight function $w(x_1,x_2)$ is determined by imposing the Hermiticity conditions,
\begin{equation}
    N_i^\dagger = N_i, \qquad i=1,2.
\end{equation}
A direct computation with the use of~\eqref{eq:adjoint-rule} yields the form
\begin{align}
    w(x_1,x_2) = &\ \frac{\left(\alpha _2+2 N-2 x_1-2 x_2+1\right) \left(\alpha _1+\alpha _2+2 N-2 x_1+2\right)}{\left(\alpha _1+\alpha _2+2 N-x_1+1\right) \left(\alpha _1+\alpha _2+2 N-x_1+2\right)}
    \nonumber\\
    &\ \times \frac{\Gamma (N+1)^2}{\Gamma \left(x_1+1\right) \Gamma \left(2 N-x_1+\alpha _1+\alpha _2+1\right)}\frac{\Gamma \left(x_2+\alpha _1+1\right)}{\Gamma \left(x_2+1\right)}
    \nonumber\\
    &\ \times \frac{\Gamma \left(N-x_1-x_2+\alpha _2+1\right){}^2}{\Gamma \left(N-x_1-x_2+1\right){}^2}\frac{\Gamma \left(2 N-2 x_1-x_2+\alpha _1+\alpha _2+2\right)}{\Gamma \left(2 N-2 x_1-x_2+\alpha _2+2\right)}.
    \label{eq:weight-function}
\end{align}
\begin{rem}
The restrictions $\alpha_i>-1$
guarantee that the weight function is strictly positive
throughout $\mathfrak{R}(N)$.
\end{rem}
In Section~\ref{sec:continuum-limit}, we will show that, under an appropriate scaling of the lattice variables and a gauge transformation, the Hamiltonian~\eqref{eq:hamiltonian-operator} converges to the continuous Smorodinsky--Winternitz I Hamiltonian.

\section{Superintegrability and exact solvability}
\label{sec:supint-exact-solv}

\subsection{Dynamical algebra}
\label{subsec:dynamical-algebra}

We assume the existence of two pairs of ladder operators $\{a_i, a_i^\dagger\}$ through which their defining action is specified through the commutation relations
\begin{equation}
    [N_i,a_j] = -\delta_{ij}\,a_j, \qquad [N_i,a_j^\dagger] = \delta_{ij}\,a_j^\dagger.
    \label{eq:Ni-aj-commutation}
\end{equation}
These relations ensure that the operators $a_i$ and $a_i^\dagger$ lower and raise, respectively, the eigenvalues of $N_i$ by one unit. This ladder structure provides a natural starting point for the construction of the dynamical algebra of the model. We also assume that each annihilation operator $a_i$ can be expressed as a linear combination of finite-difference operators,
\begin{equation}
    a_i = \sum_{|\eta_1|+|\eta_2| \le 2N} a_i^{\eta_1,\eta_2}\, \Delta_{\eta_1,\eta_2},
    \label{eq:ak-ansatz}
\end{equation}
where the coefficients $a_i^{\eta_1,\eta_2}$ are functions on $\mathfrak{R}(N)$ that vanish outside the region. Using the explicit representations of the number operators~\eqref{eq:number-operators},
the commutation relations~\eqref{eq:Ni-aj-commutation} translate into an inhomogeneous
linear system for the coefficients $a_i^{\eta_1,\eta_2}$ whose solution is subjected to the confinement on
$\mathfrak{R}(N)$, to the lowest admissible order of finite-differences and to a normalization constant.

These annihilation operators take the factorized form
\begin{align}
    &a_1 = -\frac{1}{N}a_{13}(x_1,x_2;\alpha_1,\alpha_2,N-1)a_{12}(x_1,x_2;\alpha_1+1,\alpha_2,N-1)a_{11}(x_1,x_2;\alpha_1,\alpha_2,N),
    \nonumber\\
    &a_2 = -\frac{1}{N}a_{23}(x_1,x_2;\alpha_1,\alpha_2,N-1)a_{22}(x_1,x_2;\alpha_1,\alpha_2+1,N-1)a_{21}(x_1,x_2;\alpha_1,\alpha_2,N),
    \label{eq:annihilation-operators}
\end{align}
where for $i = 1, 2$ and $j = 1, 2, 3$,
\begin{equation}
    a_{ij} = \sum_{\vec{\eta} \in S_{ij}}a_{ij}^{\eta_1,\eta_2}\, \Delta_{\eta_1,\eta_2}.
    \label{eq:contiguity-operators}
\end{equation}

For conciseness, the explicit expressions of the coefficients are deferred to Appendix~\ref{app:ladder-coefficients}. Each elementary factor $a_{ij}$ is a contiguity operator \cite{Cra-Mor-Vin-Zai-25} that shifts the parameters $(\alpha_1,\alpha_2,N)$
as displayed by its arguments in~\eqref{eq:annihilation-operators}; the full composition lowers the excitation number $n_i$ by one unit while preserving the lattice size $N$. Using the Hermitian adjoint rule~\eqref{eq:adjoint-rule}, we also find for the creation operators $a_i^\dagger$,
\begin{align}
    &a_1^\dagger = -\frac{1}{N}\bar{a}_{13}(x_1,x_2;\alpha_1,\alpha_2,N+1)\bar{a}_{12}(x_1,x_2;\alpha_1-1,\alpha_2,N+1)\bar{a}_{11}(x_1,x_2;\alpha_1,\alpha_2,N),
    \nonumber\\
    &a_2^\dagger = -\frac{1}{N}\bar{a}_{23}(x_1,x_2;\alpha_1,\alpha_2,N+1)\bar{a}_{22}(x_1,x_2;\alpha_1,\alpha_2-1,N+1)\bar{a}_{21}(x_1,x_2;\alpha_1,\alpha_2,N),
    \label{eq:creation-operators}
\end{align}
where for $i = 1, 2$ and $j = 1, 2, 3$,
\begin{equation}
    \bar{a}_{ij} = \sum_{\vec{\eta} \in \bar{S}_{ij}}\bar{a}_{ij}^{\eta_1,\eta_2}\, \Delta_{\eta_1,\eta_2}.
    \label{eq:contiguity-operators-bar}
\end{equation}
The explicit expressions of the coefficients are again deferred to Appendix~\ref{app:ladder-coefficients}. From these realizations, the mode-dependent structure functions $a_ia_j^\dagger$ and $a_i^\dagger a_j$ can be computed explicitly. They are found to be
\begin{align}
    &a_i^\dagger a_i = \frac{(N-N_1-N_2+1)^2(N_i+\alpha_i)N_i}{N^2},
    \nonumber\\
    &a_ia_i^\dagger = \frac{(N-N_1-N_2)^2(N_i+\alpha_i+1)(N_i+1)}{N^2},
    \nonumber\\
    &(N-N_1-N_2)^2a_i^\dagger a_j = (N-N_1-N_2+1)^2a_ja_i^\dagger, \qquad i \neq j.
    \label{eq:structure-function}
\end{align}
The complete algebraic structure of the \emph{dynamical algebra} can thus be summarized as
\begin{align}
    &[N_i,N_j] = [a_i,a_j] = 0,
    \qquad
    [N_i,a_j] = -\delta_{ij}a_j,
    \qquad
    [a_i,a_i^\dagger] = a_ia_i^\dagger - a_i^\dagger a_i,
    \nonumber\\
    &(N-N_1-N_2+1)^2[a_i, a_j^\dagger] = -(2(N-N_1-N_2)+1)a_j^\dagger a_i, \qquad i \neq j.
    \label{eq:dynamical-algebra}
\end{align}
These relations provide the algebraic framework underlying the discrete model and will be used in the next subsections to construct its symmetry algebra and solve the spectral problem.

\subsection{Symmetry algebra}

We now exhibit the symmetries of the Hamiltonian~\eqref{eq:hamiltonian-operator}. In accordance with the continuous Smorodinsky--Winternitz I model as will be shown in Section \ref{sec:continuum-limit}, we consider the following algebraically independent operators,
\begin{equation}
    C_1 = H_1,
    \qquad
    C_2 = \{a_1,a_2^\dagger\} + \{a_1^\dagger,a_2\} - 4H_1H_2,
    \label{eq:symmetry-operators}
\end{equation}
where $\{\cdot,\cdot\}$ denotes the anti-commutator. Using the commutation relations of the dynamical algebra \eqref{eq:dynamical-algebra}, it follows directly that
\begin{equation}
    [H,C_i] = 0,
\end{equation}
and thus, $C_i$ and $H$ are symmetry operators. Since the model admits three integrals of motion (by including the Hamiltonian) in two dimensions, it follows that the discrete Smorodinsky--Winternitz I system is \textit{maximally superintegrable}. A direct computation shows that $C_i$ obey the following cubic algebra,
\begin{align}
    &[C_1,C_2] \equiv C_3,
    \nonumber\\
    &[C_1,C_3] = \alpha C_1^2 + \gamma C_1 + \delta C_2,
    \nonumber\\
    &[C_2,C_3] = \mu C_1^3 + \nu C_1^2 - \alpha\{C_1,C_2\} + \xi C_1 - \gamma C_2 + \zeta.
    \label{eq:cubic-algebra}
\end{align}
The constants which characterize the corresponding algebra are given by
\begin{align}
    &\alpha = -4, \qquad \gamma = 4H, \qquad \delta = 1, \qquad \mu = 8\left(\eta^2-4\right), \qquad \nu = -12\left(\eta^2-4\right)H,
    \nonumber\\
    &\xi = \eta^2\left(\left(1-\alpha_1^2\right)+\left(1-\alpha_2^2\right)\right) + 4\left(\eta^2-4\right)H^2, \qquad \zeta = -\eta^2\left(1-\alpha_1^2\right)H
    \nonumber\\
    &\eta = \frac{(N-N_1-N_2)^2+(N-N_1-N_2+1)^2}{N^2}.
\end{align}
\begin{rem}
The quantities $\gamma,\mu,\nu,\xi,\zeta$ are structure
functions rather than scalar constants: each is central, being a
function of the Hamiltonian $H$ (and, through $\eta$, of
$N_1+N_2 = H-\tfrac{\alpha_1+\alpha_2}{2}-1$), and therefore commutes with all
the generators $C_i$.
\end{rem}
We shall refer to the algebra generated by
$\{C_1,C_2,C_3,H\}$ as the discrete two-dimensional
Smorodinsky--Winternitz I algebra,
denoted by $\mathcal{SW}^N_I(2)$. It admits the Casimir operator of the generalized Daskaloyannis type \cite{Das-91,Das-01,Mar-09},
\begin{equation}
    C = C_3^2 - \alpha\{C_1^2,C_2\} - \gamma\{C_1,C_2\} - \delta C_2^2 + \tfrac{\mu}{2}C_1^4 + \tfrac{2\nu}{3}C_1^3 + (\tfrac{\delta\mu}{2}+\alpha^2+\xi)C_1^2 + (\tfrac{\delta\nu}{3}+\alpha\gamma+2\zeta)C_1.
\end{equation}
A direct computation shows that in the present realization, $C$ takes the form
\begin{equation}
    C = -\frac{\eta^2}{4}\left(1-\alpha_1^2\right)\left(4H^2+\left(1-\alpha_2^2\right)\right).
\end{equation}
It is of interest to note that under the transformations
\begin{equation}
    \hat{C}_1 = C_1, \quad \hat{C}_2 = - 2(2-\eta)C_1^2 + 2(2-\eta)HC_1 + C_2,
    \label{eq:trans-symmetry-operators}
\end{equation}
we recover the Hahn algebra \cite{Fra-Gab-Vin-Vin-Zhe-19,granovskii1992mutual}:
\begin{align}
    &[\hat{C}_1,\hat{C}_2] \equiv \hat{C}_3,
    \nonumber\\
    &[\hat{C}_2,\hat{C}_3] = 2\eta\{\hat{C}_1,\hat{C}_2\} - 2\eta H\hat{C}_2 + \eta^2\left(\left(1-\alpha_1^2\right)+\left(1-\alpha_2^2\right)\right)\hat{C}_1 - \eta^2\left(1-\alpha_1^2\right)H,
    \nonumber\\
    &[\hat{C}_3,\hat{C}_1] = 2\eta\hat{C}_1^2 - 2\eta H\hat{C}_1 - \hat{C}_2.
    \label{eq:hahn-algebra-sw}
\end{align}

The emergence of the Hahn algebra therefore provides an algebraic explanation for the appearance of dual Hahn polynomials in the spectral theory of the model as will be shown in the next subsection. In this sense, the symmetry algebra and the exact polynomial solutions constitute two complementary manifestations of the same underlying structure.

\subsection{Exact solutions}
\label{subsec:exact-solutions}

Having shown the model to be maximally superintegrable, we now turn to its exact solvability, in agreement with the conjecture proposed by Tempesta, Turbiner and Winternitz \cite{Tem-Tur-Win-00}. Since the number operators \eqref{eq:number-operators} commute with each other, there exists a set of simultaneous orthonormal energy eigenstates $\{\ket{n_1,n_2}\}$ such that
\begin{align}
    &N_i\ket{n_1,n_2} = n_i\ket{n_1,n_2},
    \label{eq:number-action}
    \\
    &\braket{m_1,m_2|n_1,n_2} = \delta_{m_1,n_1}\delta_{m_2,n_2}.
    \label{eq:ortho-rel-energy-ket}
\end{align}
From the mode-dependent structure functions \eqref{eq:structure-function}, it follows that the energy spectrum is bounded from below and from above,
\begin{align}
    \begin{rcases}
    a_i\ket{n_1,n_2} = 0 \qquad \text{if} \qquad n_i = 0
    \nonumber\\
    a_i^\dagger\ket{n_1,n_2} = 0 \qquad \text{if} \qquad n_1 + n_2 = N
    \end{rcases}
    \implies (n_1,n_2) \in \mathfrak{R}(N).
\end{align}
The action of the ladder operators then reads as
\begin{align}
    &a_1\ket{n_1,n_2} = \frac{N-n_1-n_2+1}{N}\sqrt{n_1(n_1+\alpha_1)}\ket{n_1-1,n_2},
    \nonumber\\
    &a_1^\dagger\ket{n_1,n_2} = \frac{N-n_1-n_2}{N}\sqrt{(n_1+1)(n_1+\alpha_1+1)}\ket{n_1+1,n_2},
    \nonumber\\
    &a_2\ket{n_1,n_2} = \frac{N-n_1-n_2+1}{N}\sqrt{n_2(n_2+\alpha_2)}\ket{n_1,n_2-1},
    \nonumber\\
    &a_2^\dagger\ket{n_1,n_2} = \frac{N-n_1-n_2}{N}\sqrt{(n_2+1)(n_2+\alpha_2+1)}\ket{n_1,n_2+1},
    \label{eq:ladder-action}
\end{align}
where we have chosen the coefficients to be real and positive by convention. By expanding the energy eigenstate with the use of \eqref{eq:state-expansion},
\begin{equation}
    \ket{n_1,n_2} = \sum_{(x_1,x_2) \in \mathfrak{R}(N)} w(x_1,x_2)P_{n_1,n_2}(x_1,x_2)\ket{x_1,x_2}, \qquad \braket{x_1,x_2|n_1,n_2} = P_{n_1,n_2}(x_1,x_2),
\end{equation}
the corresponding orthogonality relation follows from \eqref{eq:ortho-rel-energy-ket} and \eqref{eq:ortho-rel-position},
\begin{equation}
    \sum_{(x_1,x_2) \in \mathfrak{R}(N)}w(x_1,x_2)P_{m_1,m_2}^*(x_1,x_2)P_{n_1,n_2}(x_1,x_2) = \delta_{m_1,n_1}\delta_{m_2,n_2},
    \label{eq:ortho-rel-energy-fct}
\end{equation}
while the corresponding difference equations follow from \eqref{eq:number-action} and \eqref{eq:number-operators},
\begin{equation}
    N_iP_{n_1,n_2}(x_1,x_2) = n_iP_{n_1,n_2}(x_1,x_2), \qquad i = 1,2.
    \label{eq:diff-eq-energy}
\end{equation}

Referring to \cite{geronimo2010bispectrality,tratnik1991some}, the solutions satisfying the orthogonality relation \eqref{eq:ortho-rel-energy-fct} and the difference equations \eqref{eq:diff-eq-energy} have been studied in details and they are given by the dual Hahn limit of the bivariate Racah polynomials of Tratnik type after an appropriate change of variables, (see Appendix \ref{app:Tratnik-poly} for more details)
\begin{align}
    P_{n_1,n_2}(x_1,x_2) = & \ p_{n_1}(x_1,x_2;\alpha_1,\alpha_2,N)q_{n_1,n_2}(x_1,x_2;\alpha_2,N),
    \label{eq:energy-eigenfct}
    \\
    p_{n_1}(x_1,x_2;\alpha_1,\alpha_2,N) = & \ (-1)^{n_1}\frac{d_{n_1}(x_1;x_1+x_2-\alpha_2-2N-2,x_1+x_2-\alpha_1-\alpha_2-2N-2,-x_1-x_2-1)}{(-N)_{n_1}\sqrt{\Gamma(n_1+1)\Gamma(n_1+\alpha_1+1)}}
    \nonumber\\
    q_{n_1,n_2}(x_1,x_2;\alpha_2,N) = & \ (-1)^{n_2}\frac{d_{n_2}(x_1+x_2-n_1;n_1-N-1,n_1-\alpha_2-N-1,n_1-N-1)}{(n_1-N)_{n_2}\sqrt{\Gamma(n_2+1)\Gamma(n_2+\alpha_2+1)}},
    \label{eq:sep-energy-eigenfct}
\end{align}
where
\begin{equation}
    d_n(x;\alpha,\delta,\gamma) = (\alpha+1)_n(\gamma+1)_n\hat{d}_n(\lambda(x);\gamma,\delta,-\alpha-1),
    \label{eq:mod-dual-Hahn}
\end{equation}
with $\hat{d}_{n}$ being the dual Hahn polynomials defined in \eqref{eq:dual-Hahn}. The properties of the energy eigenfunctions are then summarized in the following way,
\begin{align}
    &HP_{n_1,n_2} = (n_1+n_2+\tfrac{\alpha_1+\alpha_2}{2}+1)P_{n_1,n_2},
    \nonumber\\
    &a_1 P_{n_1,n_2} = \frac{N-n_1-n_2+1}{N}\sqrt{n_1(n_1+\alpha_1)}P_{n_1-1,n_2},
    \nonumber\\
    &a_1^\dagger P_{n_1,n_2} = \frac{N-n_1-n_2}{N}\sqrt{(n_1+1)(n_1+\alpha_1+1)}P_{n_1+1,n_2},
    \nonumber\\
    &a_2 P_{n_1,n_2} = \frac{N-n_1-n_2+1}{N}\sqrt{n_2(n_2+\alpha_2)}P_{n_1,n_2-1},
    \nonumber\\
    &a_2^\dagger P_{n_1,n_2} = \frac{N-n_1-n_2}{N}\sqrt{(n_2+1)(n_2+\alpha_2+1)}P_{n_1,n_2+1}.
    \label{eq:prop-eigenfct}
\end{align}
This completes the algebraic solution of the discrete Smorodinsky--Winternitz I model. The spectrum, ladder-operator structure and eigenfunctions are thus obtained explicitly.

\section{Continuum limit}
\label{sec:continuum-limit}

The continuum limit provides a stringent consistency check for the construction of the discrete model. Indeed, if it is to be regarded as a genuine finite discrete realization of the continuous Smorodinsky--Winternitz I system, it should reproduce not only the continuous Hamiltonian, but also its ladder operators, its eigenfunctions and its symmetry algebra.

We introduce a rescaled set of variables,
\begin{equation}
    x_1(x,y) = N - \sqrt{N(x+y)},
    \qquad
    x_2(x,y) = N - \sqrt{Ny} - x_1(x,y),
    \label{eq:change-of-variables}
\end{equation}
where, in the limit $N \to \infty$, one has $x,y \ge 0$. Equivalently,
\begin{equation}
    x(x_1,x_2)=\frac{(N-x_1)^2-(N-x_1-x_2)^2}{N},
    \qquad
    y(x_1,x_2)=\frac{(N-x_1-x_2)^2}{N}.
    \label{eq:inverse-change-of-variables}
\end{equation}
Under the elementary shifts
\[
(x_1,x_2)\mapsto (x_1+a,x_2+b),
\qquad a,b\in\mathbb{Z},
\]
the induced variations of the continuum variables $x(x_1,x_2)$ and $y(x_1,x_2)$ are respectively,
\begin{equation}
    \delta x(a,b)
    =
    \frac{a^2}{N}
    -2a\sqrt{\frac{x+y}{N}}-\delta y(a,b),
    \qquad
    \delta y(a,b)
    =
    \frac{(a+b)^2}{N}
    -2(a+b)\sqrt{\frac{y}{N}}.
    \label{eq:delta-continuum}
\end{equation}
Thus, for a function $f$, the finite difference operator \eqref{eq:difference-operator} can be rewritten as a Taylor expansion  around the induced variations $\delta x$ and $\delta y$:
\begin{equation}
    \Delta_{a,b} f(x_1,x_2)
    =
    f(x_1+a,x_2+b)-f(x_1,x_2)
    =
    f(x+\delta x,y+\delta y)-f(x,y)
    =
    \sum_{\substack{i,j\geq 0\\(i,j)\neq(0,0)}}
    \frac{\delta x^i}{i!}
    \frac{\delta y^j}{j!}
    \partial_x^i\partial_y^j f(x,y) .
    \label{eq:taylor-difference}
\end{equation}
Substituting \eqref{eq:change-of-variables} and using
\eqref{eq:delta-continuum}--\eqref{eq:taylor-difference} on the number operators \eqref{eq:number-operators} and the ladder operators \eqref{eq:annihilation-operators} and \eqref{eq:creation-operators} before taking $N \to \infty$, one obtains
the following limits:
\begin{align}
    \lim_{N\to\infty}N_1
    &=
    -x\partial_x^2-(\alpha_1+1-x)\partial_x,
    \nonumber\\
    \lim_{N\to\infty}N_2
    &=
    -y\partial_y^2-(\alpha_2+1-y)\partial_y,
    \nonumber\\
    \lim_{N\to\infty}a_1
    &=
    -x\partial_x^2-(\alpha_1+1)\partial_x,
    \nonumber\\
    \lim_{N\to\infty}a_2
    &=
    -y\partial_y^2-(\alpha_2+1)\partial_y,
    \nonumber\\
    \lim_{N\to\infty}a_1^\dagger
    &=
    -x\partial_x^2-(\alpha_1+1-2x)\partial_x+\alpha_1+1-x,
    \nonumber\\
    \lim_{N\to\infty}a_2^\dagger
    &=
    -y\partial_y^2-(\alpha_2+1-2y)\partial_y+\alpha_2+1-y.
    \label{eq:limit-operators}
\end{align}
These operators are the differential, forward-shift and
backward-shift operators associated with the normalized Laguerre polynomials; see
\eqref{eq:Laguerre-diff} and \eqref{eq:Laguerre-shift}. It is therefore natural to expect that the eigenfunctions solved in Section~\ref{subsec:exact-solutions} converge to a product of normalized Laguerre polynomials.

Indeed, using the hypergeometric identity \eqref{eq:identity-hypergeometric} and the limit relation \eqref{eq:limit-hypergeometric} in \eqref{eq:sep-energy-eigenfct}, one finds
\begin{align}
    \lim_{N\to\infty}
    p_{n_1}\bigl(x_1(x,y),x_2(x,y);\alpha_1,\alpha_2,N\bigr)
    &=
    \frac{L_{n_1}^{(\alpha_1)}(x)}
    {\sqrt{\Gamma(n_1+\alpha_1+1)/n_1!}},
    \nonumber\\
    \lim_{N\to\infty}
    q_{n_1,n_2}\bigl(x_1(x,y),x_2(x,y);\alpha_2,N\bigr)
    &=
    \frac{L_{n_2}^{(\alpha_2)}(y)}
    {\sqrt{\Gamma(n_2+\alpha_2+1)/n_2!}},
    \nonumber\\
    \lim_{N\to\infty}P_{n_1,n_2}\bigl(x_1(x,y),x_2(x,y)\bigr)
    &=\frac{L_{n_1}^{(\alpha_1)}(x)}
    {\sqrt{\Gamma(n_1+\alpha_1+1)/n_1!}}\frac{L_{n_2}^{(\alpha_2)}(y)}
    {\sqrt{\Gamma(n_2+\alpha_2+1)/n_2!}}.
    \label{eq:dual-Hahn-to-Laguerre}
\end{align}
Thus, the bivariate dual Hahn polynomials of Tratnik type reduce to a product of normalized Laguerre polynomials.

Let us next consider the weight function. Stirling's formula applied to
\eqref{eq:weight-function} gives, under the scaling
\eqref{eq:change-of-variables} and in the limit $N\gg1$,
\begin{equation}
    w\bigl(x_1(x,y),x_2(x,y)\bigr)
    \approx
    \frac{4\sqrt{(x+y)y}}{N}
    x^{\alpha_1}y^{\alpha_2}e^{-x-y}.
    \label{eq:weight-continuum-limit}
\end{equation}
Moreover, the determinant of the Jacobian matrix yields
\begin{equation}
    \left|
    \frac{\partial(x_1,x_2)}{\partial(x,y)}
    \right|
    =
    \frac{N}{4\sqrt{(x+y)y}}.
    \label{eq:jacobian-continuum}
\end{equation}
Consequently,
\begin{equation}
    \left|
    \frac{\partial(x_1,x_2)}{\partial(x,y)}
    \right|
    w\bigl(x_1(x,y),x_2(x,y)\bigr)
    \xrightarrow{N\to\infty}
    x^{\alpha_1}y^{\alpha_2}e^{-x-y}.
\end{equation}
In the continuum limit, the orthogonality relation \eqref{eq:ortho-rel-energy-fct} therefore
converges to
\begin{align}
    &\int_0^\infty\int_0^\infty
    x^{\alpha_1}y^{\alpha_2}e^{-x-y}
    \frac{
    L_{m_1}^{(\alpha_1)}(x)L_{n_1}^{(\alpha_1)}(x)
    L_{m_2}^{(\alpha_2)}(y)L_{n_2}^{(\alpha_2)}(y)}
    {
    \sqrt{
    \frac{\Gamma(m_1+\alpha_1+1)}{m_1!}
    \frac{\Gamma(n_1+\alpha_1+1)}{n_1!}
    \frac{\Gamma(m_2+\alpha_2+1)}{m_2!}
    \frac{\Gamma(n_2+\alpha_2+1)}{n_2!}
    }}
    \,dx\,dy = \delta_{m_1,n_1}\delta_{m_2,n_2}.
    \label{eq:ortho-rel-cont}
\end{align}
This is precisely the orthogonality relation of the Laguerre polynomials \eqref{eq:Laguerre-orthogonality}. To recover the standard Schrödinger realization of the continuous Smorodinsky--Winternitz I system, we now rescale
\[
x\mapsto \frac{x^2}{2},
\qquad
y\mapsto \frac{y^2}{2},
\]
and perform a gauge transformation with
\begin{equation}
    g(x,y)
    =
    x^{\alpha_1+1/2}y^{\alpha_2+1/2}
    e^{-(x^2+y^2)/4}.
    \label{eq:SWI-gauge-factor}
\end{equation}
Proceeding with the limiting operators \eqref{eq:limit-operators}, one obtains
\begin{align}
    \cA_1
    &=
    g\left(\left.\lim_{N\to\infty}a_1\right|_{x\mapsto x^2/2}\right)g^{-1}
    =
    -\frac12\left(
    \partial_x^2+x\partial_x+\frac{x^2}{4}
    +\frac{1-4\alpha_1^2}{4x^2}
    +\frac12
    \right),
    \nonumber\\
    \cA_1^\dagger
    &=
    g\left(\left.\lim_{N\to\infty}a_1^\dagger\right|_{x\mapsto x^2/2}\right)g^{-1}
    =
    -\frac12\left(
    \partial_x^2-x\partial_x+\frac{x^2}{4}
    +\frac{1-4\alpha_1^2}{4x^2}
    -\frac12
    \right),
    \nonumber\\
    \cN_1
    &=
    g\left(\left.\lim_{N\to\infty}N_1\right|_{x\mapsto x^2/2}\right)g^{-1}
    =
    -\frac12\partial_x^2
    +\frac{x^2}{8}
    -\frac{1-4\alpha_1^2}{8x^2}
    -\frac{\alpha_1+1}{2},
    \nonumber\\
    \cA_2
    &=
    g\left(\left.\lim_{N\to\infty}a_2\right|_{y\mapsto y^2/2}\right)g^{-1}
    =
    -\frac12\left(
    \partial_y^2+y\partial_y+\frac{y^2}{4}
    +\frac{1-4\alpha_2^2}{4y^2}
    +\frac12
    \right),
    \nonumber\\
    \cA_2^\dagger
    &=
    g\left(\left.\lim_{N\to\infty}a_2^\dagger\right|_{y\mapsto y^2/2}\right)g^{-1}
    =
    -\frac12\left(
    \partial_y^2-y\partial_y+\frac{y^2}{4}
    +\frac{1-4\alpha_2^2}{4y^2}
    -\frac12
    \right),
    \nonumber\\
    \cN_2
    &=
    g\left(\left.\lim_{N\to\infty}N_2\right|_{y\mapsto y^2/2}\right)g^{-1}
    =
    -\frac12\partial_y^2
    +\frac{y^2}{8}
    -\frac{1-4\alpha_2^2}{8y^2}
    -\frac{\alpha_2+1}{2}.
    \label{eq:continuous-ladder-operators}
\end{align}
It follows that
\begin{align}
    \cH
    &=
    \cN_1+\cN_2+\frac{\alpha_1+\alpha_2}{2}+1
    \nonumber\\
    &=
    -\frac12\partial_x^2-\frac12\partial_y^2
    +\frac18(x^2+y^2)
    -\frac{1-4\alpha_1^2}{8x^2}
    -\frac{1-4\alpha_2^2}{8y^2}.
    \label{eq:continuous-SWI-Hamiltonian}
\end{align}
This is the continuous Smorodinsky--Winternitz I system as defined in \cite{Mil-Pos-Win-13}. Its associated normalized eigenfunctions are \cite{Tem-Tur-Win-00}
\begin{align}
    \Psi_{n_1,n_2}(x,y)
    &=
    g(x,y)
    \lim_{N\to\infty}
    P_{n_1,n_2}
    \bigl(
    x_1(x^2/2,y^2/2),
    x_2(x^2/2,y^2/2)
    \bigr)
    \nonumber\\
    &=
    \sqrt{\frac{n_1!}{\Gamma(\alpha_1+n_1+1)}}
    x^{\alpha_1+1/2}e^{-x^2/4}
    L_{n_1}^{(\alpha_1)}(x^2/2)
    \nonumber\\
    &\qquad\times
    \sqrt{\frac{n_2!}{\Gamma(\alpha_2+n_2+1)}}
    y^{\alpha_2+1/2}e^{-y^2/4}
    L_{n_2}^{(\alpha_2)}(y^2/2).
    \label{eq:SWI-continuum-wavefunctions}
\end{align}
It remains to examine the symmetry algebra. On the one hand, from an analytic perspective, applying the same limiting procedure to \eqref{eq:trans-symmetry-operators} gives:
\begin{align}
    \cC_1
    &=
    g\left(\left.\lim_{N\to\infty}C_1\right|_{x\mapsto x^2/2}\right)g^{-1}
    =
    -\frac12\partial_x^2+\frac{x^2}{8}
    -\frac{1-4\alpha_1^2}{8x^2},
    \nonumber\\
    \cC_2
    &=
    g\left(
    \left.\lim_{N\to\infty}C_2\right|_{x\mapsto x^2/2,\;y\mapsto y^2/2}
    \right)g^{-1}
    =
    \frac12(x\partial_y-y\partial_x)^2
    +\frac{(1-4\alpha_1^2)y^2}{8x^2}
    +\frac{(1-4\alpha_2^2)x^2}{8y^2}
    -\frac14.
    \label{eq:continuous-SWI-symmetries}
\end{align}
These operators satisfy the following commutation relations,
\begin{align}
    [\cC_1,\cC_2]
    &\equiv
    \cC_3,
    \nonumber\\
    [\cC_2,\cC_3]
    &=
    4\{\cC_1,\cC_2\}
    -4\cH\cC_2
    +4\bigl((1-\alpha_1^2)+(1-\alpha_2^2)\bigr)\cC_1
    -4(1-\alpha_1^2)\cH,
    \nonumber\\
    [\cC_3,\cC_1]
    &=
    4\cC_1^2
    -4\cH\cC_1
    -\cC_2.
    \label{eq:continuous-SWI-algebra}
\end{align}
This is the continuous two-dimensional Smorodinsky--Winternitz I algebra as given in \cite{Mil-Pos-Win-13}, denoted by $\cS\cW_I(2)$. On the other hand, from an algebraic perspective, it is readily seen that the discrete two-dimensional Smorodinsky--Winternitz I algebra \eqref{eq:hahn-algebra-sw}, in the limit $N \to \infty$, converges to the continuous two-dimensional Smorodinsky--Winternitz I algebra \eqref{eq:continuous-SWI-algebra},
\begin{equation}
    \cS\cW_I^N(2) \xrightarrow{N\to\infty}\cS\cW_I(2).
\end{equation}
\begin{rem}
    Both $\cS\cW_I^N(2)$ and $\cS\cW_I(2)$ have a Hahn algebra presentation. This is a consequence of the structure function $\eta$ tending to a finite non-zero value, as $N \to \infty$; the companion model \cite{Ber-Che-Vin-26} behaves differently in this respect.
\end{rem}
Having recovered the continuous Hamiltonian and its eigenfunctions, and the symmetry algebra, the discrete model may therefore be viewed as a genuine finite realization of the continuous model rather than merely a finite-difference approximation of its equations of motion.

\section{Conclusion}
\label{sec:conc}

We have introduced a finite discrete analogue of the
Smorodinsky--Winternitz I system defined on a triangular region of the two-dimensional square lattice. The model is constructed from a pair of commuting number operators and a set of ladder operators generating its dynamical algebra. We have shown that the resulting system is maximally superintegrable and that its symmetry algebra admits a Hahn-algebra presentation.

The spectral problem has been solved exactly in terms of the bivariate dual Hahn polynomials of Tratnik type. These functions provide an orthogonal basis of energy eigenfunctions and realize irreducible representations of the symmetry algebra. This establishes a direct link between multivariate orthogonal polynomials and finite superintegrability.

The continuum limit has been analyzed in detail. After an appropriate scaling of the lattice variables and a gauge transformation, the model reduces to the continuous Smorodinsky--Winternitz I system together with its associated symmetry algebra. The present construction therefore extends the family of finite superintegrable models that retain the principal algebraic and spectral features of their continuous counterparts.

Together with the companion discretization of the Smorodinsky--Winternitz II system \cite{Ber-Che-Vin-26}, the present construction suggests that finite lattice realizations preserving maximal superintegrability are considerably more widespread than previously recognized. This opens the possibility of constructing discrete counterparts of other superintegrable systems while retaining their algebraic, spectral and symmetry structures.

\section*{Acknowledgments}
VVC benefits from a scholarship from the FRQNT. The work of LV is supported in part through a Natural Sciences and Engineering Research Council (NSERC) of Canada.

\section*{Conflict of interest}
The authors state that there is no conflict of interest.

\section*{Data availability}
This manuscript has no associated data.

\appendix

\section{Coefficients of the ladder operators}
\label{app:ladder-coefficients}

The coefficients of the annihilation operators \eqref{eq:annihilation-operators},\eqref{eq:contiguity-operators} are given by
\begin{align}
    S_{11} &= \{(1,0),(0,1)\},
    \nonumber\\
    a_{11}^{1,0} &= -\frac{1}{\alpha _1+\alpha _2+2 N-2 x_1+1} = -a_{11}^{0,1},
    \nonumber\\ \nonumber\\
    S_{12} &= \{(1,-1),(0,0)\},
    \nonumber\\
    a_{12}^{1,-1} &= -\frac{x_2 \left(\alpha _2+2 N-2 x_1-x_2+1\right)}{\alpha _1+\alpha _2+2 N-2 x_1+1}, \qquad a_{12}^{0,0} = \alpha_1,
    \nonumber\\ \nonumber\\
    S_{13} &=\{(0,0),(-1,1),(-2,2),(-1,0),(0,-1),(-2,1),(-1,-1),(-2,0),(0,-2)\},
    \nonumber\\
    a_{13}^{0,0} &= (N+1)^2,
    \nonumber\\
    a_{13}^{-1,1} &= \frac{2 x_1 \left(N-x_1-x_2+1\right){}^2 \left(\alpha _1+x_2+1\right) \left(\alpha _2+2 N-2 x_1-x_2+3\right) \left(\alpha _1+\alpha _2+2 N-x_1+4\right)}{\left(\alpha _2+2 N-2 x_1-2 x_2+2\right)_2 \left(\alpha _1+\alpha _2+2 N-2 x_1+3\right) \left(\alpha _1+\alpha _2+2 N-2 x_1+5\right)},
    \nonumber\\
    a_{13}^{-2,2} &= \frac{\left(x_1-1\right)_2 \left(N-x_1-x_2+1\right){}^2 \left(\alpha _1+x_2+1\right)_2}{\left(\alpha _2+2 N-2 x_1-2 x_2+2\right)_2 \left(\alpha _1+\alpha
   _2+2 N-2 x_1+4\right)_2},
   \nonumber\\
   a_{13}^{-1,0} &= \frac{x_1 \left(\alpha _1+\alpha _2+2 N-x_1+4\right) \left(\alpha _2 \left(2 N-2 x_1-2 x_2+3\right)+2 \left(N-x_1-x_2+1\right)_2\right)}{\left(\alpha _2+2 N-2 x_1-2 x_2+2\right) \left(\alpha _2+2 N-2
   x_1-2 x_2+4\right)}
   \nonumber\\
   &\quad\times \left(\frac{\left(x_2+1\right) \left(\alpha _1+x_2+1\right)}{\left(\alpha _1+\alpha _2+2 N-2 x_1+4\right)_2}+\frac{\left(\alpha _2+2 N-2 x_1-x_2+3\right) \left(\alpha _1+\alpha _2+2 N-2
   x_1-x_2+3\right)}{\left(\alpha _1+\alpha _2+2 N-2 x_1+3\right)_2}\right)
   \nonumber\\
   a_{13}^{0,-1} &= \left(\alpha _2 \left(2 N-2 x_1-2 x_2+3\right)+2 \left(N-x_1-x_2+1\right)_2\right)
   \nonumber\\
   &\quad \times \frac{x_2 \left(\alpha _2+2 N-2 x_1-x_2+3\right) \left(\alpha _1+\alpha _2+2 N-x_1+3\right)_2}{\left(\alpha _2+2 N-2 x_1-2 x_2+2\right) \left(\alpha _2+2 N-2 x_1-2 x_2+4\right) \left(\alpha _1+\alpha _2+2 N-2 x_1+3\right)_2},
   \nonumber\\
   a_{13}^{-2,1} &= \left(\alpha _2 \left(2 N-2 x_1-2 x_2+3\right)+2 \left(N-x_1-x_2+1\right)_2\right)
   \nonumber\\
   &\quad \times \frac{\left(x_1-1\right)_2 \left(\alpha _1+x_2+1\right) \left(\alpha _1+\alpha _2+2 N-2 x_1-x_2+4\right)}{\left(\alpha _2+2 N-2 x_1-2 x_2+2\right) \left(\alpha _2+2 N-2 x_1-2 x_2+4\right) \left(\alpha _1+\alpha _2+2 N-2 x_1+4\right)_2},
   \nonumber\\
   a_{13}^{-1,-1} &= \frac{2 x_1 x_2 \left(\alpha _2+N-x_1-x_2+2\right){}^2 \left(\alpha _1+\alpha _2+2 N-x_1+4\right) \left(\alpha _1+\alpha _2+2 N-2 x_1-x_2+4\right)}{\left(\alpha _2+2 N-2 x_1-2 x_2+3\right)_2 \left(\alpha _1+\alpha _2+2 N-2 x_1+3\right) \left(\alpha _1+\alpha _2+2 N-2 x_1+5\right)},
   \nonumber\\
   a_{13}^{-2,0} &= \frac{\left(x_1-1\right)_2 \left(\alpha _2+N-x_1-x_2+2\right){}^2 \left(\alpha _1+\alpha _2+2 N-2 x_1-x_2+4\right)_2}{\left(\alpha _2+2 N-2 x_1-2 x_2+3\right)_2 \left(\alpha _1+\alpha _2+2 N-2 x_1+4\right)_2},
   \nonumber\\
   a_{13}^{0,-2} &= \frac{\left(x_2-1\right)_2 \left(\alpha _2+N-x_1-x_2+2\right){}^2 \left(\alpha _1+\alpha _2+2 N-x_1+3\right)_2}{\left(\alpha _2+2 N-2 x_1-2 x_2+3\right)_2 \left(\alpha _1+\alpha _2+2 N-2 x_1+3\right)_2},
    \label{eq:a1-coefficients}
\end{align}
\begin{align}
    S_{21} =&\ \{(1,0),(0,1),(1,-1)\},
    \nonumber\\
    a_{21}^{1,0} =& -\frac{\alpha _2+2 N-2 x_1-x_2}{\left(\alpha _2+2 N-2 x_1-2 x_2\right) \left(\alpha _1+\alpha _2+2 N-2 x_1+1\right)},
    \nonumber\\
    a_{21}^{0,1} =& -\frac{\alpha _1+x_2+1}{\left(\alpha _2+2 N-2 x_1-2 x_2\right) \left(\alpha _1+\alpha _2+2 N-2 x_1+1\right)},
    \nonumber\\
    a_{21}^{1,-1} =&\ \frac{x_2}{\left(\alpha _2+2 N-2 x_1-2 x_2\right) \left(\alpha _1+\alpha _2+2 N-2 x_1+1\right)},
    \nonumber\\ \nonumber\\
    S_{22} =&\ \{(1,0),(0,1),(0,0),(1,-1)\},
    \nonumber\\
    a_{22}^{1,0} =& -\frac{\left(N-x_1-x_2\right){}^2 \left(\alpha _2+2 N-2 x_1-x_2\right)}{\left(\alpha _2+2 N-2 x_1-2 x_2\right) \left(\alpha _1+\alpha _2+2 N-2 x_1+1\right)},
    \nonumber\\
    a_{22}^{0,1} =& -\frac{\left(N-x_1-x_2\right){}^2 \left(\alpha _1+x_2+1\right)}{\left(\alpha _2+2 N-2 x_1-2 x_2\right) \left(\alpha _1+\alpha _2+2 N-2 x_1+1\right)},
    \nonumber\\
    a_{22}^{0,0} =&\ \alpha_2,
    \qquad
    a_{22}^{1,-1} =\ \frac{x_2 \left(\alpha _2+N-x_1-x_2\right){}^2}{\left(\alpha _2+2 N-2 x_1-2 x_2\right) \left(\alpha _1+\alpha _2+2 N-2 x_1+1\right)},
    \nonumber\\ \nonumber\\
    S_{23} =&\ S_{13}, \qquad a_{23}^{\eta_1,\eta_2} = a_{13}^{\eta_1,\eta_2}.
    \label{eq:a2-coefficients}
\end{align}
The coefficients of the creation operators \eqref{eq:creation-operators},\eqref{eq:contiguity-operators-bar} are given by
\begin{align}
    \bar{S}_{11} =&\ \{(0,0),(-1,0),(0,-1)\},
    \nonumber\\
    \bar{a}_{11}^{0,0} =& \left(x_1+x_2\right) \left(\alpha _2+2 N-x_1-x_2+3\right)-x_1 \left(\alpha _1+\alpha _2+2 N-x_1+3\right),
    \nonumber\\
    \bar{a}_{11}^{-1,0} =& -\frac{x_1 \left(\alpha _1+x_2\right) \left(\alpha _1+\alpha _2+2 N-2 x_1-x_2+3\right)}{\alpha _1+\alpha _2+2 N-2 x_1+3},
    \nonumber\\
    \bar{a}_{11}^{0,-1} =&\ \frac{x_2 \left(\alpha _2+2 N-2 x_1-x_2+3\right) \left(\alpha _1+\alpha _2+2 N-x_1+3\right)}{\alpha _1+\alpha _2+2 N-2 x_1+3},
    \nonumber\\ \nonumber\\
    \bar{S}_{12} =&\ \{(0,0),(-1,1)\},
    \nonumber\\
    \bar{a}_{12}^{0,0} =&\ 1, \qquad \bar{a}_{12}^{-1,1} = -\frac{x_1}{\alpha _1+\alpha _2+2 N-2 x_1+3},
    \nonumber\\ \nonumber\\
    \bar{S}_{13} =&\ \{(2,0),(0,2),(1,1),(2,-1),(1,0),(0,1),(2,-2),(1,-1),(0,0)\},
    \nonumber\\
    \bar{a}_{13}^{2,0} =&\ \frac{\left(-N+x_1+x_2+1\right){}^2 \left(\alpha _2+2 N-2 x_1-x_2-2\right)_2}{\left(\alpha _2+2 N-2 x_1-2 x_2-2\right)_2 \left(\alpha _1+\alpha
   _2+2 N-2 x_1-1\right)_2},
   \nonumber\\
    \bar{a}_{13}^{0,2} =&\ \frac{\left(-N+x_1+x_2+1\right){}^2 \left(\alpha _1+x_2+1\right)_2}{\left(\alpha _2+2 N-2 x_1-2 x_2-2\right)_2 \left(\alpha _1+\alpha _2+2 N-2 x_1\right)_2},
    \nonumber\\
    \bar{a}_{13}^{1,1} =&\ \frac{2 \left(-N+x_1+x_2+1\right){}^2 \left(\alpha _1+x_2+1\right) \left(\alpha _2+2 N-2 x_1-x_2-1\right)}{\left(\alpha _2+2 N-2 x_1-2 x_2-2\right)_2 \left(\alpha _1+\alpha _2+2 N-2
   x_1-1\right) \left(\alpha _1+\alpha _2+2 N-2 x_1+1\right)},
   \nonumber\\
   \bar{a}_{13}^{2,-1} =&\ \frac{x_2 \left(\alpha _2+2 N-2 x_1-x_2-1\right) \left(\alpha _2 \left(2 N-2 x_1-2 x_2-1\right)+2 \left(N-x_1-x_2-1\right)_2\right)}{\left(\alpha _2+2 N-2 x_1-2 x_2-2\right) \left(\alpha _2+2 N-2 x_1-2
   x_2\right) \left(\alpha _1+\alpha _2+2 N-2 x_1-1\right)_2},
   \nonumber\\
   \bar{a}_{13}^{1,0} =&\ \frac{\alpha _2 \left(2 N-2 x_1-2 x_2-1\right)+2 \left(N-x_1-x_2-1\right)_2}{\left(\alpha _2+2 N-2 x_1-2 x_2-2\right) \left(\alpha _2+2 N-2 x_1-2 x_2\right)}
   \nonumber\\
   &\times \left(\frac{x_2 \left(\alpha _1+x_2\right)}{\left(\alpha _1+\alpha _2+2 N-2 x_1-1\right)_2}+\frac{\left(\alpha _2+2 N-2 x_1-x_2\right) \left(\alpha _1+\alpha _2+2 N-2
   x_1-x_2\right)}{\left(\alpha _1+\alpha _2+2 N-2 x_1\right)_2}\right),
   \nonumber\\
   \bar{a}_{13}^{0,1} =&\ \frac{\left(\alpha _1+x_2+1\right) \left(\alpha _1+\alpha _2+2 N-2 x_1-x_2\right) \left(\alpha _2 \left(2 N-2 x_1-2 x_2-1\right)+2 \left(N-x_1-x_2-1\right)_2\right)}{\left(\alpha _2+2 N-2 x_1-2
   x_2-2\right) \left(\alpha _2+2 N-2 x_1-2 x_2\right) \left(\alpha _1+\alpha _2+2 N-2 x_1\right)_2},
   \nonumber\\
   \bar{a}_{13}^{2,-2} =&\ \frac{\left(x_2-1\right)_2 \left(\alpha _2+N-x_1-x_2\right){}^2}{\left(\alpha _2+2 N-2 x_1-2 x_2-1\right)_2 \left(\alpha _1+\alpha _2+2 N-2 x_1-1\right)_2},
   \nonumber\\
   \bar{a}_{13}^{1,-1} =&\ \frac{2 x_2 \left(\alpha _2+N-x_1-x_2\right){}^2 \left(\alpha _1+\alpha _2+2 N-2 x_1-x_2\right)}{\left(\alpha _2+2 N-2 x_1-2 x_2-1\right)_2 \left(\alpha _1+\alpha _2+2 N-2 x_1-1\right)
   \left(\alpha _1+\alpha _2+2 N-2 x_1+1\right)},
   \nonumber\\
   \bar{a}_{13}^{0,0} =&\ 1,
    \label{eq:a1d-coefficients}
\end{align}
\begin{align}
    \bar{S}_{21} =&\ \{(0,0),(-1,1),(-1,0),(0,-1)\},
    \nonumber\\
    \bar{a}_{21}^{0,0} =&\ \left(N-x_1-x_2+1\right){}^2-\alpha _2 \left(x_1+x_2\right),
    \nonumber\\
    \bar{a}_{21}^{-1,1} =&\ \frac{x_1 \left(N-x_1-x_2+1\right){}^2 \left(\alpha _1+x_2+1\right)}{\left(\alpha _2+2 N-2 x_1-2 x_2+2\right) \left(\alpha _1+\alpha _2+2 N-2 x_1+3\right)},
    \nonumber\\
    \bar{a}_{21}^{-1,0} =& -\frac{x_1 \left(\alpha _2+N-x_1-x_2+1\right){}^2 \left(\alpha _1+\alpha _2+2 N-2 x_1-x_2+3\right)}{\left(\alpha _2+2 N-2 x_1-2 x_2+2\right) \left(\alpha _1+\alpha _2+2 N-2 x_1+3\right)},
    \nonumber\\
    \bar{a}_{21}^{0,-1} =& -\frac{x_2 \left(\alpha _2+N-x_1-x_2+1\right){}^2 \left(\alpha _1+\alpha _2+2 N-x_1+3\right)}{\left(\alpha _2+2 N-2 x_1-2 x_2+2\right) \left(\alpha _1+\alpha _2+2 N-2 x_1+3\right)},
    \nonumber\\ \nonumber\\
    \bar{S}_{22} =&\ \bar{S}_{21},
    \nonumber\\
    \bar{a}_{22}^{0,0} =&\ 1, \qquad \bar{a}_{22}^{-1,1} = \frac{x_1 \left(\alpha _1+x_2+1\right)}{\left(\alpha _2+2 N-2 x_1-2 x_2+2\right) \left(\alpha _1+\alpha _2+2 N-2 x_1+3\right)},
    \nonumber\\
    \bar{a}_{22}^{-1,0} =& -\frac{x_1 \left(\alpha _1+\alpha _2+2 N-2 x_1-x_2+3\right)}{\left(\alpha _2+2 N-2 x_1-2 x_2+2\right) \left(\alpha _1+\alpha _2+2 N-2 x_1+3\right)},
    \nonumber\\
    \bar{a}_{22}^{0,-1} =& -\frac{x_2 \left(\alpha _1+\alpha _2+2 N-x_1+3\right)}{\left(\alpha _2+2 N-2 x_1-2 x_2+2\right) \left(\alpha _1+\alpha _2+2 N-2 x_1+3\right)},
    \nonumber\\ \nonumber\\
    \bar{S}_{23} =&\ \bar{S}_{13}, \qquad \bar{a}_{23}^{\eta_1,\eta_2} = \bar{a}_{13}^{\eta_1,\eta_2}.
    \label{eq:a2d-coefficients}
\end{align}

\section{Hypergeometric polynomials}
\label{app:hyp-poly}

\subsection{Hypergeometric function}

The hypergeometric function is defined by \cite{Koe-Les-Swa-10}
\begin{equation}
    _r{F}_s\left(\genfrac{}{}{0pt}{}{a_1,\dotsc, a_r}{b_1,\dotsc,b_s};z\right) = \sum_{k=0}^\infty \frac{(a_1)_k\cdots(a_r)_k}{(b_1)_k\cdots(b_s)_k}\frac{z^k}{k!}.
    \label{eq:hypergeometric-function}
\end{equation}
They have the limit relations
\begin{align}
    &\lim_{\lambda\to\infty} \ _r{F}_s\left(\genfrac{}{}{0pt}{}{a_1,\dotsc, a_{r-1},\lambda a_r}{b_1,\dotsc,b_s};\frac{z}{\lambda}\right) = \ _{r-1}{F}_s\left(\genfrac{}{}{0pt}{}{a_1,\dotsc, a_{r-1}}{b_1,\dotsc,b_s};a_rz\right),
    \nonumber\\
    &\lim_{\lambda\to\infty} \ _r{F}_s\left(\genfrac{}{}{0pt}{}{a_1,\dotsc,a_r}{b_1,\dotsc,b_{s-1},\lambda b_s};\lambda z\right) = \ _r{F}_{s-1}\left(\genfrac{}{}{0pt}{}{a_1,\dotsc, a_r}{b_1,\dotsc,b_{s-1}};\frac{z}{b_s}\right),
    \label{eq:limit-hypergeometric}
\end{align}
which generalizes to
\begin{align}
    \lim_{\lambda\to\infty} &\ _r{F}_s\left({a_1,\dotsc,a_{r-i},\lambda^{k_1}a_{r-i+1},\dotsc,\lambda^{k_i} a_r\atop b_1,\dotsc,b_{s-i},\lambda^{l_1}b_{s-j+1},\dotsc,\lambda^{l_j}b_s};\frac{\lambda^{l_1}\cdots\lambda^{l_j}}{\lambda^{k_1}\cdots\lambda^{k_i}}z\right)
    \nonumber\\
    &=\ _{r-i}{F}_{s-j}\left({a_1,\dotsc, a_{r-i}\atop b_1,\dotsc,b_{s-j}};\frac{a_{r-i+1}\cdots a_r}{b_{s-j+1}\cdots b_s}z\right).
    \label{eq:general-limit-hypergeometric}
\end{align}
A specific functional identity for $_3{F}_2$ is given by
\begin{align}
    &_3{F}_2\left({a_1,a_2,a_3\atop b_1,b_2};1\right) = \frac{\Gamma(b_1)\Gamma(b_1+b_2-a_1-a_2-a_3)}{\Gamma(b_1-a_1)\Gamma(b_1+b_2-a_2-a_3)} \ _3{F}_2\left({a_1,b_2-a_2,b_2-a_3\atop b_1+b_2-a_2-a_3,b_2};1\right),
    \label{eq:identity-hypergeometric}
\end{align}
where $\text{Re}(b_1+b_2-a_1-a_2-a_3),\text{Re}(b_1-a_1)>0$ and $\Gamma(z) = \int_0^\infty t^{z-1}e^{-t}dt$, $\text{Re}(z)>0$, is the Gamma function.

\subsection{Formulas for the Laguerre polynomials}

The Laguerre polynomials are defined by \cite{Koe-Les-Swa-10}
\begin{equation}
    L_n^{(\alpha)}(x) = \frac{(\alpha+1)_n}{n!}\ _1{F}_1\left({-n\atop \alpha+1};x\right).
    \label{eq:Laguerre}
\end{equation}
They satisfy the differential equation
\begin{equation}
    xL^{(\alpha)''}_n(x) + (\alpha+1-x)L^{(\alpha)'}_n(x) + nL^{(\alpha)}_n(x) = 0.
    \label{eq:Laguerre-diff}
\end{equation}
They obey the orthogonality relation
\begin{equation}
    \int_0^\infty dx \ e^{-x}x^\alpha L_m^{(\alpha)}(x)L_n^{(\alpha)}(x) = \frac{\Gamma(n+\alpha+1)}{n!}\delta_{mn}, \quad \alpha>-1.
    \label{eq:Laguerre-orthogonality}
\end{equation}
The forward and backward shift operators are respectively as follow
\begin{align}
    &-xL^{(\alpha)''}_n(x) - (\alpha+1)L^{(\alpha)'}_n(x) = (n+\alpha)L^{(\alpha)}_{n-1}(x),
    \nonumber\\
    &-xL^{(\alpha)''}_n(x) - (\alpha+1-2x)L^{(\alpha)'}_n(x) + (\alpha+1-x)L^{(\alpha)}_n(x) = (n+1)L^{(\alpha)}_{n+1}(x).
    \label{eq:Laguerre-shift}
\end{align}

\subsection{Formulas for the dual Hahn polynomials}

The dual Hahn polynomials are defined by \cite{Koe-Les-Swa-10}
\begin{equation}
    \hat{d}_n(\lambda(x);\gamma,\delta,N) = \ _3{F}_2\left({-n,-x,x+\gamma+\delta+1\atop \gamma+1,-N};1\right), \qquad \lambda(x) = x(x+\gamma+\delta+1),
    \label{eq:dual-Hahn}
\end{equation}
where $N \in \mathbb{N}$, $n = 0,1,\dotsc,N$ and $\gamma,\delta>-1$ or $\gamma,\delta<-N$.

\subsection{Formulas for the Racah polynomials}

The Racah polynomials are defined by \cite{Koe-Les-Swa-10}

\begin{align}
    &\hat{r}_n(\lambda(x);\alpha,\beta,\gamma,\delta) = \ _4{F}_3\left({-n,n+\alpha+\beta+1,-x,x+\gamma+\delta+1\atop \alpha+1,\beta+\delta+1,\gamma+1};1\right), \qquad \lambda(x) = x(x+\gamma+\delta+1)
    \label{eq:Racah}
\end{align}
where $N \in \mathbb{N}$, $n = 0,1,\dotsc,N$ and $\alpha+1 = -N$ or $\beta+\delta+1 = -N$ or $\gamma+1 = -N$.

They are related to the dual Hahn polynomials by the limits
\begin{align}
    &\lim_{\alpha \to \infty} \hat{r}_n(\lambda(x);\alpha,-\delta-N-1,\gamma,\delta) = \hat{d}_n(\lambda(x);\gamma,\delta,N),
    \nonumber\\
    &\lim_{\beta \to \infty} \hat{r}_n(\lambda(x);-N-1,\beta,\gamma,\delta) = \hat{d}_n(\lambda(x);\gamma,\delta,N),
    \nonumber\\
    &\lim_{\beta \to \infty} \hat{r}_n(\lambda(x);\alpha,\beta,-N-1,\alpha+\delta+N+1) = \hat{d}_n(\lambda(x);\alpha,\delta,N).
\end{align}

\section{Bivariate polynomials of Tratnik type}
\label{app:Tratnik-poly}

\subsection{Racah}

The bivariate polynomials of the Racah kind are given by \cite{tratnik1991some, geronimo2010bispectrality}
\begin{align}
    R_2(n;x;\beta;N) =&\ r_{n_1}(\beta_1-\beta_0-1,\beta_2-\beta_1-1,-x_2-1,\beta_1+x_2;x_1)
    \nonumber\\
    &\times r_{n_2}(2n_1+\beta_2-\beta_0-1,\beta_3-\beta_2-1,n_1-N-1,n_1+\beta_2+N;-n_1+x_2),
    \label{eq:poly-R2}
\end{align}
with $n = (n_1,n_2) \in \mathbb{N}_0$ such that $n_1+n_2 \le N \in \mathbb{N}$, $x = (x_1,x_2) \in \mathbb{R}^2$, $\beta = (\beta_0,\beta_1,\beta_2,\beta_3) \in \mathbb{R}^4$ and
\begin{equation}
    r_i(a,b,c,d;x) = (a+1)_i(b+d+1)_i(c+1)_i \ \hat{r}_i(\lambda(x);a,b,c,d),
\end{equation}
where $\hat{r}_i$ are the Racah polynomials \eqref{eq:Racah}. They satisfy the difference equations
\begin{align}
    &\mathfrak{L}^x_i \, R_2(n;x;\beta;N) = \mu_i(n)R_2(n;x;\beta;N), \qquad i=1,2.
\end{align}
The operators $\mathfrak{L}^x_i$ can be written as follows:
\begin{align}
    &\mathfrak{L}^x_1 = C_1^{(1)}(E_{x_1}-1) + C_{-1}^{(1)}(E_{x_1}^{-1}-1),
    \nonumber\\
    &\mathfrak{L}^x_2 = \sum_{0 \ne \nu \in \{-1,0,1\}^2}C_\nu^{(2)}(E_{x_1}^{\nu_1}E_{x_2}^{\nu_2}-1),
\end{align}
where $E_{x_i}$ denotes the customary shift operator acting on functions $f$ of $x$ as
\begin{equation}
    E_{x_i}f(x) = f(x+e_i),
\end{equation}
with $\{e_1,e_2\}$ being the standard basis for $\mathbb{R}^2$. The coefficients $C_\nu$ are given by
\begin{align}
    &C_1^{(1)} = \frac{\left(x_2-x_1\right) \left(\beta _1+x_1\right) \left(-\beta _0+\beta _1+x_1\right) \left(\beta _2+x_1+x_2\right)}{\left(\beta _1+2
     x_1\right) \left(\beta _1+2 x_1+1\right)},
    \nonumber\\
    &C_{-1}^{(1)} = -\frac{x_1 \left(\beta _0+x_1\right) \left(\beta _1+x_1+x_2\right) \left(\beta _1-\beta _2+x_1-x_2\right)}{\left(\beta _1+2 x_1-1\right)
    \left(\beta _1+2 x_1\right)},
    \nonumber\\
    &C_{(1,1)}^{(2)} = \frac{\left(N-x_2\right) \left(\beta _1+x_1\right) \left(-\beta _0+\beta _1+x_1\right) \left(\beta _2+x_1+x_2\right) \left(\beta
    _2+x_1+x_2+1\right) \left(\beta _3+N+x_2\right)}{\left(\beta _1+2 x_1\right) \left(\beta _1+2 x_1+1\right) \left(\beta _2+2 x_2\right)
    \left(\beta _2+2 x_2+1\right)},
    \nonumber\\
    &C_{(1,0)}^{(2)} = \frac{\left(x_2-x_1\right) \left(\beta _1+x_1\right) \left(-\beta _0+\beta _1+x_1\right) \left(\beta _2+x_1+x_2\right) \left(\left(\beta
   _2+1\right) \left(\beta _3-1\right)+2 N \left(\beta _3+N\right)+2 x_2 \left(\beta _2+x_2\right)\right)}{\left(\beta _1+2 x_1\right)
   \left(\beta _1+2 x_1+1\right) \left(\beta _2+2 x_2-1\right) \left(\beta _2+2 x_2+1\right)},
   \nonumber\\
   &C_{(0,1)}^{(2)} = \frac{\left(N-x_2\right) \left(\left(\beta _0+1\right) \left(\beta _1-1\right)+2 x_1 \left(\beta _1+x_1\right)\right) \left(\beta
   _2+x_1+x_2\right) \left(-\beta _1+\beta _2-x_1+x_2\right) \left(\beta _3+N+x_2\right)}{\left(\beta _1+2 x_1-1\right) \left(\beta _1+2
   x_1+1\right) \left(\beta _2+2 x_2\right) \left(\beta _2+2 x_2+1\right)},
   \nonumber\\
   &C_{(1,-1)}^{(2)} = \frac{\left(x_1-x_2\right) \left(x_1-x_2+1\right) \left(\beta _1+x_1\right) \left(-\beta _0+\beta _1+x_1\right) \left(\beta _2+N+x_2\right)
   \left(-\beta _2+\beta _3+N-x_2\right)}{\left(\beta _1+2 x_1\right) \left(\beta _1+2 x_1+1\right) \left(\beta _2+2 x_2-1\right) \left(\beta
   _2+2 x_2\right)},
   \nonumber\\
   &C_{(-1,1)}^{(2)} = \frac{x_1 \left(N-x_2\right) \left(\beta _0+x_1\right) \left(\beta _1-\beta _2+x_1-x_2-1\right) \left(\beta _1-\beta _2+x_1-x_2\right)
   \left(\beta _3+N+x_2\right)}{\left(\beta _1+2 x_1-1\right) \left(\beta _1+2 x_1\right) \left(\beta _2+2 x_2\right) \left(\beta _2+2
   x_2+1\right)},
   \nonumber\\
   &C_{(0,-1)}^{(2)} = \frac{\left(x_1-x_2\right) \left(\beta _1+x_1+x_2\right) \left(\left(\beta _0+1\right) \left(\beta _1-1\right)+2 \beta _1 x_1+2 x_1^2\right)
   \left(\beta _2+N+x_2\right) \left(\beta _2-\beta _3-N+x_2\right)}{\left(\beta _1+2 x_1-1\right) \left(\beta _1+2 x_1+1\right) \left(\beta
   _2+2 x_2-1\right) \left(\beta _2+2 x_2\right)},
   \nonumber\\
   &C_{(-1,0)}^{(2)} = -\frac{x_1 \left(\beta _0+x_1\right) \left(\beta _1+x_1+x_2\right) \left(\beta _1-\beta _2+x_1-x_2\right)}{\left(\beta _1+2 x_1-1\right) \left(\beta _1+2 x_1\right) \left(\beta _2+2
   x_2-1\right) \left(\beta _2+2 x_2+1\right)}
   \nonumber\\
   &\qquad\qquad\, \times \left(-\beta _2+2 N^2+\beta _3
   \left(\beta _2+2 N+1\right)+2 \beta _2 x_2+2 x_2^2-1\right),
   \nonumber\\
   &C_{(-1,-1)}^{(2)} = \frac{x_1 \left(\beta _0+x_1\right) \left(\beta _1+x_1+x_2-1\right) \left(\beta _1+x_1+x_2\right) \left(\beta _2+N+x_2\right) \left(-\beta
   _2+\beta _3+N-x_2\right)}{\left(\beta _1+2 x_1-1\right) \left(\beta _1+2 x_1\right) \left(\beta _2+2 x_2-1\right) \left(\beta _2+2
   x_2\right)}.
\end{align}
and the corresponding eigenvalues $\mu_i$ are given by
\begin{align}
    &\mu_1(n) = -n_1(n_1-1+\beta_2-\beta_0),
    \nonumber\\
    &\mu_2(n) = -(n_1+n_2)(n_1+n_2-1+\beta_3-\beta_0).
\end{align}
The bivariate polynomials of the Racah kind obey the orthogonality relation \cite{tratnik1991some}:
\begin{align}
    \sum_{x_2=0}^N\sum_{x_1=0}^{x_2}\rho(x)R_2(m;x)R_2(n;x) = \delta_{m_1,n_1}\delta_{m_2,n_2}\lambda(n),
\end{align}
where the weight $\rho(x)$ and the normalization constant $\lambda(n)$ are given by
\begin{align}
    &\rho(x) = \frac{(\beta_1)_{x_1}}{x_1!}\frac{(\varepsilon+1)_{x_1}}{(\beta_0 + 1)_{x_1}} \times \frac{\Gamma(x_2+\beta_2-x_1-\beta_1)}{(x_2-x_1)!}\frac{\Gamma(\beta_2+x_2+x_1)}{\Gamma(\beta_1+x_2+x_1+1)}\frac{(\beta_1/2+1)_{x_1}}{(\beta_1/2)_{x_1}}
    \nonumber\\
    &\qquad\qquad \times \frac{(\beta_2/2+1)_{x_2}}{(\beta_2/2)_{x_2}}\frac{(\beta_3+N)_{x_2}}{(\beta_2-\beta_3-N+1)_{x_2}}\frac{(-N)_{x_2}}{(\beta_2+N+1)_{x_2}},
    \label{eq:weight-R2}
    \\
    &\lambda(n) = (\beta_1)^{-1}n_1!\Gamma(n_1+\beta_2-\beta_1)(n_1+\beta_2-\beta_0-1)_{n_1}\frac{\Gamma(n_1+\beta_1-\beta_0)}{\Gamma(2n_1+\beta_2-\beta_0)}
    \nonumber\\
    &\quad\quad\quad\ \times (\beta_2)^{-1}n_2!\Gamma(n_2+\beta_3-\beta_2)(2n_1+n_2+\beta_3-\beta_0-1)_{n_2}\frac{\Gamma(2n_1+n_2+\beta_2-\beta_0)}{\Gamma(2n_1+2n_2+\beta_3-\beta_0)}
    \nonumber\\
    &\quad\quad\quad\ \times \frac{\Gamma(\beta_0+1)\Gamma(\beta_2+N+1)\Gamma(n_1+n_2-\beta_0+\beta_3+N)}{\Gamma(\beta_1)\Gamma(\beta_1-\beta_0)\Gamma(\beta_3-\beta_2+N)\Gamma(\beta_0+N+1)}
    \nonumber\\
    &\quad\quad\quad\ \times (\beta_3+N)_{n_1+n_2}(-N)_{n_1+n_2}(-\beta_0-N)_{n_1+n_2},
    \label{eq:norm-R2}
\end{align}
with $\varepsilon \in \mathbb{R}$.

\subsubsection{Limit relation: Racah to dual Hahn}

Under the change of variables
\begin{align}
    &\beta_0 = - \alpha_1 - \alpha_2 - \varepsilon - 2N - 3, \qquad \beta_1 = - \alpha_1 - \alpha_2 - 2N - 2,
    \nonumber\\
    &\beta_2 = - \alpha_2 - 2N - 1, \qquad \beta_3 = - 2N,
    \\
    &x_2 \mapsto x_1 + x_2 \implies E_{x_2}^{\nu_2} \mapsto E_{x_2}^{\nu_2-\nu_1},
\end{align}
where $\alpha=(\alpha_1,\alpha_2) \in \mathbb{R}^2$ and taking the limit $\varepsilon \to \infty$, we recover the Tratnik bivariate dual Hahn polynomials \eqref{eq:energy-eigenfct}, the weight function \eqref{eq:weight-function} and the number operators \eqref{eq:number-operators}:
\begin{align}
    &\lim_{\varepsilon \to \infty}\frac{R_2(n;x)}{\sqrt{(-1)^N\beta_1\beta_2\frac{\Gamma(\beta_1)\Gamma(N-\beta_2+\beta_3)}{\Gamma(N+\beta_2+1)}\lambda(n)}} = P_{n_1,n_2}(x_1,x_2),
    \nonumber\\
    &\lim_{\varepsilon \to \infty} (-1)^N\beta_1\beta_2\frac{\Gamma(\beta_1)\Gamma(N-\beta_2+\beta_3)}{\Gamma(N+\beta_2+1)}\rho(x) = w(x_1,x_2),
    \nonumber\\
    &\lim_{\varepsilon \to \infty} -\varepsilon^{-1}\mathfrak{L}_1^x = N_1, \qquad \lim_{\varepsilon \to \infty} -\varepsilon^{-1}(\mathfrak{L}_2^x-\mathfrak{L}_1^x) = N_2.
\end{align}

\bibliography{SWIDR}

\end{document}